\documentclass[12pt]{article}
\newcommand{\ket}[1]{{\left|#1\right\rangle}}

\def\com#1#2{\Big[#1,#2\Big ]}

\def\be{\begin{equation}}
\def\ee{\end{equation}}
\newcommand{\bea}{\begin{eqnarray}}
\newcommand{\eea}{\end{eqnarray}}

\def\ra{\rightarrow}
\def\fr{\frac}
\def\a{\alpha}
\def\b{\beta}

\def\d{\delta}
\def\e{\epsilon}

\def\l{\lambda}
\def\m{\mu}
\def\n{\nu}

\def\s{\sigma}

\def\th{\theta}

\def\d{\delta}

\def\Pd{\Phi^\dagger}

\def\p{\partial}

\def\nn{\noindent}
\def\no{\nonumber}

  \def\cL{{\cal L}}
  
\def\cP{{\cal P}}

\begin{document}
\title{{\small\hfill IMSc/2002/11/39}\\ $O(3)$ Sigma model with Hopf term on Fuzzy Sphere}
\author{ T. R. Govindarajan\thanks{trg@imsc.res.in} and E. Harikumar\thanks{hari@imsc.res.in}\\
The Institute of Mathematical Sciences\\C. I. T. Campus, Tharamani\\
Chennai, INDIA-600 113}
\maketitle
\begin{abstract}  
  We formulate the $O(3)~\s-$ model on fuzzy sphere and construct the
  Hopf term. We show that the field can be expanded in terms of the
  ladder operators of Holstein-Primakoff realisation of $SU(2)$
  algebra and the corresponding basis set can be classified into
  different topological sectors by the magnetic quantum numbers. We
  obtain topological charge $Q$ and show that $-2j\le Q \le2j$. We
  also construct BPS solitons. Using the covariantly conserved
  current, we construct the Hopf term and show that its value is $Q^2$
  as in the commutative case. We also point out the interesting
  relation of physical space to deformed $SU(2)$ algebra.
\end{abstract}
\medskip
\nn Keywords: Fuzzy sphere, $O(3)~\s-$ model, Hopf term, BPS solution,\,\\
\newpage
\section{Introduction}\label{intro}

Field theories on non-commutative spaces \cite{con} have been
extensively studied in last few years.  These studies have emanated
from attempts to understand the renormalisation program at a deeper
level or string theories \cite{wit}, quantum gravity and matrix models
\cite{mat}.  These theories present many interesting features and some
of these features are not shared by their commutative counterparts.
Aspects such as UV/IR mixing \cite{uv}, soliton and instanton
solutions \cite{sol,inst}, monopoles \cite{mono}, gauge invariance
\cite{gi}, renormalisability \cite{ss} etc have been widely
investigated (for a recent review, see \cite{mdnn}).

Study of field theories on compact, non-commutative spaces are of
interest as they provide an alternative way of regularising field
theories with finite degrees of freedom \cite{grosse,pres}.  Fuzzy
sphere ($S_{F}^2$) whose co-ordinates are non-commuting was introduced
and study of fuzzy field theory models was initiated in \cite{madore}.
Topological solutions in non-linear sigma model on fuzzy sphere were
obtained and their commutative limit were studied in \cite{sb}. Chiral
anomaly and the Fermion doubling problem in this setting were
investigated by Balachandran et al \cite{sv,apb}. Gauge theories on
fuzzy sphere have been introduced and studied in \cite{wata}. UV/IR
mixing in the case of $\Phi^4$ theory on fuzzy sphere has been
investigated in detail in \cite{uv1}.  Topological properties of
$S_{F}^2$ has been studied in \cite{gkp,ccy,grs}. Non-commutative
$CP^n$ model and its supersymmetric version have been studied
recently \cite{cpn,susycpn} and BPS solutions were obtained.

In this paper we study, fuzzy $CP^1$ model with Hopf term. $CP^n$
model \cite{bp} which shares many interesting properties of gauge
theories like asymptotic freedom has been a testing ground for many
ideas \cite{adda} underlying non-perturbative physics. $CP^1$ model (or
equivalently $O(3)$ non-linear sigma model) has topological solitons
which acquire fractional statistics when coupled to Hopf term
 \cite{wz}. These models with Hopf term have been analysed in canonical
framework and the existence of fractional charge was
verified \cite{wij}. $O(3)~\s-$ model has also been studied in relation
with many condensed matter systems \cite{frad}. It has been studied in
path integral framework and interesting interplay of topological and
Noether charges has been shown. The equivalence of $CP^1$ model with
Hopf term to Fermionic theory has also been shown \cite{polmpl,psm}.
It is interesting to study the $CP^1$ model with Hopf term in the
fuzzy sphere setting.

On the fuzzy sphere ($S_{F}^2$) the derivations are obtained by the
adjoint action of the angular momentum operators ${\cL}_{i}$. We take
${\cL}_{i}$ in the Holstein-Primakoff \cite{hp} realisation of $SU(2)$
algebra which acts on a finite dimensional Hilbert space. The
corresponding state vectors can be classified into different {\it
  topological} sectors using ${\cL}_{3}$ eigenvalues which are
integers. We have shown that the soliton charge $Q$ which is expressed
in terms of a covariantly conserved current is integer and equal to
the ${\cL}_{3}$ eigenvalue.  We also obtain the BPS soliton solution.
Using the covariantly conserved current, we construct the Fuzzy Hopf
term and demonstrate that its value is proportional to $Q^2$ as in the
commutative case.

Fuzzy sphere $S_{F}^2$ is obtained by quantising the co-ordinates of
$S^2$ \cite{madore}.  Thus the fuzzy sphere $(S_{F}^2)$ is defined by
the co-ordinates obeying
\be
[x_i,x_j]=i\l\e_{ijk}x_k,~~~~ x_i x_i = R^2.
\label{fuzs2}
\ee
where $R$ is the radius of the sphere. The fields on $S_{F}^2$ are
functions of the non-commutative co-ordinates and hence when one
expands the fields in terms of spherical harmonics $Y_{lm}$, there is
a cut off for the allowed values of $l$ $(l\le 2j=N)$. These fields
are realised as operators acting on a $(N+1)^2$ dimensional Hilbert
space. The fields on $S_{F}^2$ are now complex $(N+1)\times (N+1)$
matrices.

In the next section, we present the fuzzy sphere as a deformed Hopf
fibration and define the finite dimensional Hilbert space associated
with it \cite{ccy}. Here we also show how the fields are defined on
$S_{F}^2$. In section \ref{cp1}, first we present the $CP^1$ model on
$R\otimes{S_{F}^2}$. We then get the BPS bound and define topological
charge. We also show the BPS solutions for different topological
sectors. In section \ref{hopf}, we present the construction Hopf term
and evaluate it. In section \ref{suq2}, we point out the interesting
connection between $S_{F}^2$ and $SU_q(2)$ operators, details of which
will be taken up later. Finally in section \ref{con}, we present the
concluding remarks.

\section{Fuzzy Sphere}\label{fuzzy}

The coordinates of $S_{F}^2$ obeys Eqn. (\ref{fuzs2}) and hence they
are $(2j+1)\times (2j+1)$ dimensional matrices and can be realised by
the generators of the $2j+1$ dimensional irreducible representation of
$SU(2)$. In the above $\l=\fr{R}{\sqrt{j(j+1)}}$, where $R$ is the
radius of the sphere. From Eqn. (\ref{fuzs2}) we see that in the limit
$\l\rightarrow 0$ (i.e., $j\rightarrow \infty$), we recover the
commutative sphere. To get the non-commutative plane we should take
$j\to\infty,~R\to\infty$ such that $\fr{R}{j}=\l$ is a constant.
Continuum is obtained by taking $\l\to 0.$

We define the $S_{F}^2$ using the non-commutative analogue of the Hopf
fibration $S^3\rightarrow S^2$. For this purpose we start with the
creation and annihilation operators of two uncoupled harmonic
oscillators obeying
\be
\com{A_\a}{A_{\b}^\dagger}=\d_{\a\b},~~\a=1,2;~A_1=a, A_2=b
\ee
with $a^\dagger a = N_1$ and $b^\dagger b= N_2.$ The operators
$a,~a^\dagger,~b,~b^\dagger$ act on the infinite dimensional number
states
\be
{\ket{n}}_{1,~2}=\fr{(A_{\a}^\dagger)^n}{\sqrt{n!}}\ket{0}_{1,~2}~,~~~n=0,1,2,..
\label{neq}
\ee
with $A _{\a}{\ket{0}}_{1,~2}=0.$ Using these one can define two sets of operators
\be
J_{+}^{(\a)}= A_{\a}^\dagger~\sqrt{N-N_{\a}},~~~
J_{-}^{(\a)}=\sqrt{N-N_{\a}}~~ A_{\a},~~~~J_{3}^{(\a)}=(N_{\a}-\fr{N}{2}). \label{lad1}
\ee
Each of this (Holstein-Primakoff realisation) forms a spin
$\fr{N}{2}=j$ representation of a $SU(2)$ for any given value of $N$.
For maintaining the unitarity of the representation of $SU(2),$ these
ladder operators in Eqn. (\ref{lad1}) should act only on a finite
dimensional subspaces $n, m\le N.$ Here, we note that in the case of
$SU_q(2)$, action of ladder operators can be restricted on a finite
dimensional space in a natural way by taking $q$ to be an $(N+1)th$
root of unity (see section \ref{suq2}).

Using the operators$~a,~a^\dagger,~b,~b^\dagger$, we define
\bea
{\cL}_{+} &=& a^\dagger~\sqrt{N-a^\dagger a} ~+~ b^\dagger~\sqrt{N-b^\dagger b},\no\\
{\cL}_{-} &=& \sqrt{N-a^\dagger a}~~ a ~+~\sqrt{N-b^\dagger b}~~b,\label{lad0}\\
{\cL}_{3} &=& (a^\dagger a + b^\dagger b -N).\label{hpr}
\eea
These operators obey
\be
\com{{\cL}_{+}}{{\cL}_{-}}= 2{\cL}_{3},~~~~~
\com{{\cL_{3}}}{{\cL}_{\pm}}=\pm {\cL}_{\pm}.
\ee
The states of Hilbert space on which these ladder operators ${\cL}_{\pm}$ act are\\
\begin{center}$
\begin{array}{ccc}
\nonumber
\ket{N,\ N}&m=~N&\\ \nonumber
\ldots&\\ \nonumber
\ldots\ldots\ldots&\\ \nonumber
\ket{N,\ 1}~~\ket{N-1,\ 2}~....\ket{2,\ N-1}~~\ket{1,\ N}&m=~1&\\ \nonumber
\ket{N,\ 0}~~\ket{N-1,\ 1}~..............~\ket{1,\ N-1}~~\ket{0,\ N}&m=~0&\\ \nonumber
\ket{N-1,\ 0}~~\ket{N-2,\ 1}~....\ket{1,\ N-1}~~\ket{0,\ N}&m=-1&\\ \nonumber
\ldots\ldots\ldots&\\ \nonumber
\ldots&\\ \nonumber
\ket{0,\ 0}&m=-N&
\end{array}
$
\end{center}
where $m=0,{\pm 1},..{\pm N}, N=2j$ are the corresponding ${\cL}_{3}$
{\it integer} eigenvalues which classify these states. For a given
state, $m= n_1 + n_2 -N$ where $n_1~{\rm and}~ n_2$ are the
eigenvalues of $N_1=a^\dagger a ~{\rm and}~ N_2= b^\dagger b$
respectively.  For any given $j$, there are $(2j+1)^2$ number of
states which are classified into $2N+1$ sectors by the corresponding
value of $m$. Notice that for any given value of $j~(=\fr{N}{2})$, $m$
is always integer. Note that there is a symmetry between the number of
states for any given value $n$ and $-n$ of $m.$ Any row in the above
with a given $m$ value can be thought as a representation of $SU(2)$
with spin $j=\fr{N}{2},\fr{N-1}{2},\fr{N-2}{2},...~$. Ladder
operators of this $SU(2)$ are:\\
\nn (a)~for$~ m=-N~$to$~0,$\\
$$ j_{+}=a^\dagger b,~j_{-}=b^\dagger a~{~\rm and~~}
j_3= \fr{1}{2}(a^\dagger a - b^\dagger b).$$
\nn (b) for~$m=1~$to $N$\\
\bea
J_{+}=a^\dagger \sqrt{\fr{(N-N_1)(N-N_2)}{(N_1+1)(N_2+1)}}~b,~~J_{-}= J_{+}^\dagger,~~
j_3=\fr{1}{2} (a^\dagger a - b^\dagger b).\no
\eea
A generic scalar field $\Phi$ on the fuzzy sphere is defined as
\be
\Phi =\sum_{k,l} a_{kl}(t){\cL}_{+}^k{\cL}_{-}^l,~~~k,l\le N.\label{scal}
\ee
From this we find
\be
\com{{\cL}_{3}}{\Phi}=\sum_{kl} (k-l)a_{kl}(t){\cL}_{+}^k{\cL}_{-}^l,
\ee
which for given values of $k - l$ becomes
\be
\com{{\cL}_{3}}{\Phi}= m\Phi,~~ k-l\equiv m=0,{\pm 1},..{\pm N}.
\ee
We notice from Eqn. (\ref{scal}) that under
\bea
{\cL}_{\pm}&~\rightarrow&~e^{\pm i\psi} {\cL}_{\pm},\no\\
\Phi&~\rightarrow~&\Phi e^{im\psi}.\label{gi}
\eea
We will see in the next section that this gauge transformation is a symmetry of $CP^1$ model.

Here we see that the number of independent functions $a_{kl}$ for any given $N$ is same
as obtained if one expands the scalar field in terms of the spherical harmonics, i.e.,
$\phi=\sum_{lm} A_{lm}Y_{lm}.$

In \cite{cpn} Holstein-Primakoff realisation of $SU(2)$ generators was
used to study the $CP^n$ model. Here the fields $\Phi$ are expanded in
terms of the creation and annihilation operators of the uncoupled
harmonic oscillators, $A_{\a}^\dagger,~A_\a,~ \a=1,2$. In contrast,
here we expand the fields $\Phi$ in terms of ${\cal L}_{\pm}$. In the
present case, the basis set is given by $\ket{N, 0}$, $\ket{N-1,
  1}$,$.....$ $\ket{1, N-1}$, $\ket{0, N}$ and all other states are
generated from these by the action of ${\cal L}_{\pm}.$ As pointed out
earlier, number of states in topological sectors with ${\cal L}_3$
eigenvalues $\pm m$ are same.

Differentiation on fuzzy sphere is realised by the adjoint action of
${\cL}_i$. We work with ${\cL}_i$ in anti-hermitian representation
so that $[{\cL}_i,{\cL}_j]=\e_{ijk}{\cL}_k$. Thus we have
\be
{\cL}_{i}^\dagger =-{\cL}_i,~~~{\cL}_i \Phi =[{\cL}_i,\Phi],~~({\cL}_i\Phi)^\dagger=[{\cL}_i,\Phi^\dagger]= 
{\cL}_i \Phi^\dagger.
\ee

\section{$CP^1$ model on $R\otimes{S_{F}^2}$}\label{cp1}

In this section, we present the action for $CP^1$ model on
$R\otimes{\rm fuzzy~sphere}$. Here the time coordinate which is $R$
commutes with coordinates of fuzzy sphere. This, in the planar
non-commutative limit, will lead to $CP^1$ model on
$R\otimes{R_{\th}^2}$ and in commutative limit gives $CP^1$ model in
$2+1$ dimensions.

The action for $CP^1$ model on $R\otimes{S_{F}^2}$ is given by
\be
S=\fr{2}{2j+1}tr {\int}_t \p_0\Pd \p_0\Phi +\Pd \p_0\Phi\Pd \p_0\Phi -
{[{\cL}_i,\phi]}^\dagger[{\cL}_i,\phi] -\Pd {\cL}_i\Phi \Pd {\cL}_i\Phi.
\label{cpact}
\ee
Here $\Phi$ is a non-zero complex doublet ${\rm i.e}~~\Pd =
(\Phi_{1}^*~ \Phi_{2}^*)$ which obeys the condition $\Phi^\dagger \Phi=1
\footnote{In the limit $N\rightarrow \infty,$ ${\cL}_i\ra iL_i$, we get the action
on $R\otimes{S^2}$ and by taking $\fr{L_i}{R}\ra p_i$, we get
commuting plane -$R^3$.}$. We rewrite the above action as
\be
S= Tr ~\left(\p_0\Pd \p_0\Phi +A_{0}^2 +2iA_0\Pd \p_0\Phi -
({[{\cL}_i,\phi]}^\dagger[{\cL}_i,\phi]+A_{i}^2+2iA_i \Pd {\cL}_i\Phi)\right).
\label{linact}
\ee
Here $Tr =\fr{2}{2j+1}~tr{\int}_t$.
By eliminating $A_0$ and $A_i$ using their equations of motion from
(\ref{linact}), we get back (\ref{cpact}). Above action along with
the constraint $\Pd \Phi=1$ can be expressed as
\be
S=Tr ~\left(|D_0\Phi|^2 -|D_i\Phi|^2 +\l (\Pd \Phi -1)\right).
\label{cpnl}
\ee 
Here, 
\be D_0\Phi = \p_0 \phi -i\Phi A_0,~~~D_i\Phi = {\cL}_i\Phi-i\Phi A_i.  
\ee 
It is easy to see that the equations of motion for
fields $\Phi$ as well as $\Phi^\dagger$ have equal powers of ${\cal  L}_{\pm}$. 
Thus the fields in a given topological sector stay in the same sector under time 
evolution.  The action Eqn. (\ref{cpnl}) is invariant under the right action of 
local $U(1)$ group,

\bea
\Phi&\ra&\Phi G,\label{vect1}\\
A_\m&\ra&G^\dagger A_\m G + i d_\m G^\dagger G,\label{vect}
\eea
where $G\in U$ and $d_0\equiv \p_0,~{\rm and}~ d_i\equiv {\cL}_i$.
This invariance allows us to remove the extra degree of freedom (that
of the phase of $\Phi$). From Eqn. (\ref{gi}), we identify $G=exp
~{{\cL}_3}\Phi.$ Equations of motion of $A_\m$ following from Eqn.
(\ref{cpnl}) gives
\be
A_\m= -i\Pd d_\m \Phi,\label{vec}.
\ee
We can re-write using (\ref{vec}),
\be
D_\m \Phi ={\cP}d_\m \Phi,~~~ (D_\m \Phi)^\dagger = (d_\m\Phi)^\dagger {\cP}
\ee
where ${\cP}= 1-\Phi\Pd,~{\rm and}~{\cP}^2={\cP}$. Here we note that for any $\Phi$ obeying
$\com{{\cL}_3}{\Phi}=im\Phi$,
\bea
A_3&=&-i\Pd {\cL}_3 \Phi~=~-i(\Phi_{1}^*{\cL}_3\Phi_1+\Phi_{2}^*{\cL}_3\Phi_{2})\no\\
&=&(m_1|\Phi_1|^2+m_2|\Phi_2|^2)~=~ m \Pd \Phi=m I,
\label{a3m}
\eea
where we have taken $m_1=m_2=m$.(i.e., both $\Phi_1$ and $\Phi_2$ have same
eigenvalue for ${\cL}_3$.) We have
\be
\Pd D_\m \Phi=0,\label{peqn},
\ee
and with Eqn. (\ref{a3m}) and $\com{{\cL}_i}{\Phi}=im\Phi$ we get
\be
D_3\Phi = {\cL}_3 \Phi -i\Phi A_3 =0,\label{a30}.
\ee
This Eqn. (\ref{a30}) is a consequence of the gauge invariance under Eqn. (\ref{vect1},\ref{vect}).
Using these, we define
\be
\Pd [D_\m, D_\n]\Phi=-iF_{\m\n}=~-i\left[d_\m A_\n -d_\n A_\m +i [A_\m, A_\n]-\e_{\m\n 0}A_3\right],
\label{fs}
\ee
where $F_{\m\n},~\m,\n= 0,1,2$ is the field strength on $R\otimes{S_{F}^2}.$ Notice that the last term in
(\ref{fs}), viz $-\e_{0\m\n}A_3$ is present only in the fuzzy case and it vanishes in the continuum limit.

Under the gauge transformation the field strength transform as
\be
F_{\m\n}\ra G^\dagger F_{\m\n} G
\ee
The Bianchi identity following from Eqn. (\ref{fs}),~ is
\be
\e_{\m\n\l}D_\m (\Phi F_{\n\l})=0,
\ee
which, after using Eqn. (\ref{peqn})~ leads to
\bea
&\e_{\m\n\l}D_\m F_{\m\l}=0&
\label{bian}\\
&=\e_{\m\n\l}D_\m\left(d_\n A_\l -d_\l A_\m + i[A_\n,A_\l]\right),
\label{bian1}&
\eea
where we have used the fact that $D_\m A_3=0.$

\subsection{Topological Charge}\label{topq}

We now construct lower bound on energy for the $CP^1$ model defined by
the action Eqn. (\ref{cpact}) in a topological sector and show that is
is proportional to the topological charge. In static case, with the
gauge choice $A_0=0$, we get
\be
E=\fr{1}{2j+1}tr~|D_i\Phi|^2.
\ee
Using the identity
\be
 \fr{1}{2}tr \left[ (D_i\Phi \pm i\e_{ij}D_j\Phi)^\dagger
(D_i\Phi \pm i\e_{ij}D_j\Phi)\right] +\fr{1}{2} tr (D_3\Phi)^\dagger(D_3\Phi)\ge 0,\label{bpseqn}
\ee
where $i=1,2.$ We see
\bea
\fr{1}{2j+1} tr ~|D_i\Phi|^2 +|D_3\Phi|^2&\ge& \mp \fr{i}{2j+1} tr~\e_{ij}(D_i\Phi)^\dagger D_j\Phi=Q,
\label{qchar}\\
~{\rm i.e.,}~~E&\ge& |Q|
\eea
which is re-expressed using Eqn. (\ref{fs})~ as
\be
Q=-\fr{1}{2j+1} tr~ F_{12}=-\fr{1}{2j+1} tr~ J_0,\label{qf}
\ee
where $J_0$ is the zeroeth component of
\be
J_\m=\e_{\m\n\l}F_{\n\l},
\label{cocur}
\ee
which is covariantly conserved due to Eqn. (\ref{bian}), i.e,~$D_\m J_\m=0.$

From Eqn. (\ref{qchar})~ we have
\bea
Q &=&\fr{i}{2j+1} tr~ \left [ (D_1\Phi)^\dagger (D_2\Phi)-(D_2\Phi)^\dagger(D_1\Phi)
\right]\\
&=& -\fr{i}{2j+1} tr ~ \Pd [{\cal L}_3, \Phi] =  m,
\label{qchar1}
\eea
where $\Pd[{\cL}_3,\Phi]= \Phi_{1}^*[{\cL}_3, \Phi_1] +\Phi_{2}^* [{\cL}_3, \Phi_2] =
 i( m_1|\Phi_1|^2 + m_2|\Phi_2|^2=im|\Phi|^2=imI ).$ We can also use Eqns. (\ref{qf}, \ref{fs}) and
Eqn. (\ref{a3m}) and get $Q=m.$

\subsection{BPS Solution}\label{bps}

The BPS equations following from Eqn. (\ref{bpseqn}) are
\bea
D_i\Phi \pm i\e_{ij}D_j\Phi =0,
\label{bpseq}\\
D_3\Phi =0.\label{bpse2}
\eea
$\Phi$ which satisfy the above, saturates the bound and has topological charge $Q=m.$
We recall from Eqn. (\ref{a3m}) and Eqn. (\ref{a30}) that for any $\Phi$ which is an eigenvector of ${\cL}_3$,
Eqn. (\ref{bpse2}) is identically satisfied.

Eqn. (\ref{bpseq}) can be re expressed as
\be
{\cal P}{\cL}_{\pm}\Phi=0\label{bpseq3}
\ee
where ${\cal P}=1-\Phi\Pd$ and we have used Eqn. (\ref{hpr}) and Eqn. (\ref{vec}). Thus any configuration
satisfying
\be
\com{{\cL}_{\pm}}{\Phi}=0
\ee
will be a BPS (anti-) soliton solution. Thus the generic BPS solutions  for (anti-) soliton with $Q=\pm m$ are
\be
\Phi_{soliton} =a_{m0}{\cL}_{+}^{m},~~~~~~\Phi_{anti-soliton} = a_{0m}{\cL}_{-}^{m}.
\ee
With the parameterisation
\be
\Phi={\cal F}\fr{1}{\sqrt{{\cal F}^\dagger{\cal F}}},
\ee
BPS equation (\ref{bpseq3}) becomes
\be
{\cal P}\com{{\cL}_{\pm}}{{\cal F}}\fr{1}{\sqrt{{\cal F}^\dagger {\cal F}}}=0,
\ee
where
${\cal P}=1-{\cal F}\fr{1}{\sqrt{{\cal F}^\dagger {\cal F}}} {\cal F}^\dagger$. 
It is easy to see that both $\Phi$ and ${\cal F}$ have same ${\cal L}_{3}$ 
eigenvalue. Any ${\cal F}$ satisfying
\be
\com{{\cL}_{\pm}}{{\cal F}}=0
\ee
will give a BPS (anti)-soliton solution configuration.

\section{ Hopf term}\label{hopf}

Using the covariantly conserved current $J_\m$ given in Eqn.
(\ref{cocur}), we construct the Hopf invariant term as
\be
H= \fr{1}{2\pi} Tr~\left[ \e_{\m\n\l}\left (A_\m F_{\n\l} -\fr{2i}{3}A_\m A_\n A_\l +
\e_{\n\l 0}A_\m A_3\right)\right]\label{ht}
\ee
Under the transformation $\d A_\m$ of $A_\m$,
\be
\d H = \fr{1}{\pi} Tr~ \e_{\m\n\l}\d A_\m F_{\n\l},
\ee
which is total derivative (after using Eqn. (\ref{bian1})). This sets
$H={\rm interger}.$ Using Eqn. (\ref{fs}) and Eqn. (\ref{vec}), we
express $H$ as 
\bea H&=&-\fr{1}{\pi}\left[ Tr ~\e_{\m\n\l} \Pd d_\m
  \Phi d_\n \Pd d_\l\Phi + Tr~ \Pd\p_0\Phi\Pd
  [j_3,\Phi]\right]\no\\
&=&-\fr{1}{\pi}\left[ Tr~ \e_{\m\n\l} \Pd d_\m \Phi d_\n \Pd d_\l\Phi
  +im Tr~\Pd\p_0\Phi\right]
\label{hop}
\eea
Hopf term given above, in the planar limit reduces to
\be
H=\fr{1}{2\pi}\e_{\m\n\l} \Pd \p_\m\Phi \p_\n\Pd \p_\l\Phi
\ee
with $m=0.$

For evaluating the Hopf term, we rotate static soliton configuration through $\th=2n\pi$ by
\be
e^{\th {\cal L}_3} \Phi=e^{i\th m}\Phi,~~~~~\Pd e^{\th {\cal L}_3} =\Pd e^{i\th m}\no
\ee
Under this rotation
\bea
\e_{0ij} {\cal L}_i\Pd {\cal L}_j\Phi&\ra& \e_{0ij}{\cal L}_i\Pd {\cal L}_j\Phi\no\\
\Pd\p_0\Phi &\ra& i\p_0\th~ m.
\label{calh}
\eea
Using the fact that $\int dt \p_0 \th= \th(t)-\th(0)=2\pi$(or integer multiple)
we get the Hopf term in Eqn. (\ref{hop}) to be
\be
H= m^2.
\label{hv}
\ee
From Eqn. (\ref{qchar1}) and Eqn. (\ref{hv}), we get the relation $H=Q^2$ which is exactly same as that in the
commutative plane.

\section{Fuzzy Sphere and $SU_q (2)$}\label{suq2}

The coordinates of the fuzzy sphere, which is obtained by quantising
$S^2$ obeys $SU(2)$ algebra. In section \ref{fuzzy} we have seen that
these coordinates can be realised by the generators of $SU(2)$. These
generators ${\cL}_i$ can be constructed from the creation and
annihilation operators of two uncoupled oscillators, either by
Schwinger or Holstein-Primakoff realisation of $SU(2)$. Since these
creation and annihilation operators acts on infinite dimensional
oscillator space, one has to restrict the action of ${\cL}_i$ to a
finite dimensional subspace. In section \ref{fuzzy}, we have done this
by imposing the condition $n\le N,~m\le N$ (see Eqn. (\ref{neq}) and
discussion after Eqn.(\ref{lad1})). We show here that in the case of
$SU_q(2)$ such a restriction can be introduced in a natural way and
hence the generators of $SU_q(2)$ \cite{su2q} can be used as
realisation of the coordinates of fuzzy sphere. In this section, we
briefly discuss this aspect. First, for the completeness and also to
fix the notations we present the $SU_q(2)$ construction through
q-oscillator \cite{bied}

The q-creation and annihilation operators obey
\be
a_q a_{q}^\dagger -q^{\fr{1}{2}}a_{q}^\dagger a_q =q^{-\fr{N}{2}}
\ee
where the number operator satisfy
\be
\com{N}{a_q}=-a_q,~~\com{N}{a_{q}^\dagger}=a_{q}^\dagger.
\ee
These operators act on the q-oscillator states obeying
\bea
\ket{n}_q &=& \fr{(a_{q}^\dagger)^n}{\sqrt{[n]!}} \ket{0}_q,~~N\ket{n}_q = n\ket{n},\no\\
a_q\ket{0}_q&=&0,~~~~{\rm where} ~~[n] = \fr{q^{\fr{n}{2}}-q^{-\fr{n}{2}}}{q^{\fr{1}{2}}-q^{-\fr{1}{2}}}
\equiv [n]_q.
\eea
Here by taking $q$ to be $(n+1)th$ root of unity, i.e.,
$q=exp~{\fr{2\pi i}{n+1}}$, we get $[n+1]_q=0$. With this $q$ we get
only a finite dimensional space with basis $\ket{0}_q,...,\ket{n}_q.$
Thus by fixing the value of $n$ in the definition for $q$, we are
guaranteed to have a finite dimensional Hilbert space.

With two uncoupled q-oscillator operators
\be
a_q a_{q}^\dagger -q^{\fr{1}{2}} a_{q}^\dagger a_q =q^{-\fr{N_1}{2}},~~~~
b_q b_{q}^\dagger -q^{\fr{1}{2}} b_{q}^\dagger b_q =q^{-\fr{N_2}{2}}
\ee
the Schwinger realisation of $SU_q(2)$ generators \cite{bied} are given by
\be
J_{+}= a_{q}^\dagger b_q,~~J_{-}=b_{q}^\dagger a_{q},~~J_3=\fr{1}{2}(N_1-N_2),\no
\ee
obeying
\be
\com{J_3}{J_{\pm}}=\pm J_{\pm}~~\com{J_{+}}{J_{-}}=[2J_3]_q.
\ee
These generators now acts on a finite dimensional states. The q- analogue of Holstein-Primakoff realisation of
$SU_q(2)$ generators \cite{zachos} are
\bea
{\cL}_{\pm}^{(q)}=L_{\pm}^{(1)}\otimes q^{\fr{L_{3}^{(2)}}{2}} + q^{-\fr{L_{3}^{(1)}}{2}}\otimes L_{\pm}^{(2)}
\label{qlad}\\
{\cL}_{3}^{(q)} = L_{3}^{(1)}\otimes 1 + 1 \otimes L_{3}^{(2)},
\label{ql3}
\eea
obeying
\be
\com{{\cL}_{+}^{(q)}}{{\cL}_{-}^{(q)}}= [2L_{3}^{(q)}]_q~,~~\com{L_{3}^{(q)}}{{\cL}_{\pm}^{(q)}}
= \pm{\cL}_{\pm}^{(q)},
\ee
where
\bea
L_{+}^{(1)}=a_{q}^\dagger\sqrt{N-N_1}, ~~L_{-}^{(1)}=(L_{+}^{(1)})^\dagger,
~~L_{3}^{(1)}=N_1-\fr{N}{2}\no\\
L_{+}^{(2)}=b_{q}^\dagger\sqrt{N-N_2}, ~~L_{-}^{(2)}=(L_{+}^{(2)})^\dagger,~~L_{3}^{(2)}=N_2-\fr{N}{2}
\eea
The above $q-$operators ${\cL}_{\pm}$~(Eqn.\ref{qlad}) acts exactly on the q-analogue of the states on
which the ladder operators in Eqn. (\ref{lad0}) acts. Note here that we do not have to fix any condition on $n$ and $m$ 
unlike in the case of $SU(2)$.

In terms of the $SU_q(2)$ generators Eqn. (\ref{qlad}) and Eqn. (\ref{ql3}) we can define the generic scalar field
\be
\Phi=\sum_{k,l}a_{kl}({\cL}_{+}^{(q)})^k({\cL}_{-}^{(q)})^l,
\ee
obeying
\be
\com{{\cL}_{3}^{(q)}}{\Phi}=\sum_{k.l} (k-l) a_{kl}({\cL}_{+}^{(q)})^k({\cL}_{-}^{(q)})^l.
\ee
Thus as in the section \ref{fuzzy}, with given $k$ and $l$, we get
\be
\com {{\cL}_{3}^{(q)}} {\Phi} =m \Phi, ~~k-l\equiv m.
\ee
Study of $CP^1$ model with Hopf term in this setting will be presented elsewhere.

\section{Conclusion}\label{con}

In this paper we have constructed the $CP^1$ model with Hopf term on
$R\otimes{S_{F}^2}.$ We have shown that the fields on fuzzy sphere can
be expressed in terms of the ladder operators ${\cal L}_{\pm}$ of the
Holstein-Primakoff realisation of $SU(2)$ algebra and they are
classified into different sectors by the ${\cal L}_3$ eigenvalues. The
finite dimensional Hilbert space of these operators (for a given
$N=2j$) can be classified into different sectors by $m,~-2N\le m \le
2N$ which is the eigenvalue of ${\cal L}_3.$ Moreover, number of
states with a given ${\cal L}_3$ eigenvalue $m=\pm n$ are the same.
Using the fields defined in terms of ${\cal L}_{\pm}$, we have
formulated the $CP^1$ model on $R\otimes{S_{F}^2}$ and obtained BPS
bound. The local gauge invariance sets the ${\cal L}_3$ eigenvalues of
both $CP^1$ fields $\Phi_1$ and $\Phi_2$ to be same. The current
associated with the topological charge $Q$ of the (anti-) solition is
conserved covariantly. This charge is shown to be the ${\cal L}_3$
eigenvalue of the $CP^1$ fields. We have also obtained the general
solutions for the BPS equations. For any given $j$, there are $2N+1$
distinct soliton-anti-soliton solutions. Using the covariantly
conserved current, we have then constructed the Hopf term for
$R\otimes{S_{F}^2}$ and showed that in the commutative limit, it goes
to the known form. We have also evaluated the Hopf term and showed
that it is equal to $Q^2$ as in the commutative case. We have argued
that the generators of $SU_q(2)$ allows a natural realisation of fuzzy
sphere since the dimension of the Hilbert space corresponding to
$SU_q(2)$ can be tuned to given $n$ by taking $q$ to be $n+1$th root
of unity.

It has been shown earlier that the $O(3)~\s-$ model with Hopf term
with a coefficient $\pi$ is equivalent to spin $\fr{1}{2}$ theory with
four Fermi interactions \cite{psm}. Having developed the $O(3)~\s-$
model on fuzzy sphere, one can now analyse this equivalence in this
non-perturbative formulation. It has been shown that the $CP^1$ model
with Hopf term is equivalent to spin $s$ theory. Here the spin $s$ is
related to the coefficient of the Hopf term $\theta$ and is given by
$\theta=\fr{\pi}{2s},~s= \fr{1}{2}, 1, \fr{3}{2},...$ \cite{psm}.
Studies to see whether these equivalences go through in the present
case are in progress.

\nn {\bf ACKNOWLEDGEMENTS}\\                
We thank A. P. Balachandran for usefull discussions. We also learnt
that A. P. Balachandran and Giorgio Immirzi have studied the
topological issues in $CP^1$ model in a different formalism
\cite{apbgi}.


\begin{thebibliography}{99}
\bibitem{con} A. Connes, Noncommutative geometry, Academic Press, London, (1994).
\bibitem{wit} N. Seiberg and E. Witten, JHEP {\bf 09} (1999) 032.
\bibitem{mat}A. Connes, M. R. Douglas and A. Schwarz, JHEP {\bf 9802} (1998) 033.
\bibitem{uv} S. Minwalla, M. Van Raamsdonk and N. Seiberg, JHEP {\bf 0002} (2000) 020;
A. Matusis, L. Susskind, N. Toumbas, JHEP {\bf 0012} (2000) 002; B. P. Dolan, D. O'Conner and P. Presnajder, 
JHEP {\bf 0203} (2002) 013.
\bibitem{sol} N. A. Nekrasov, 'Trieste lectures on solitons in noncommutative
gauge theories', hep-th/0011095; J. A. Harvey, 'Komaba lectures on non-commutative
solotions and D-branes', hep-th/0102076, R. Gopakumar, S. Minwalla and A. Strominger,
JHEP {\bf 0005} (2000) 020.
\bibitem{inst} N. Nekrasov and A. Schwarz, Comm. Math. Phys. {\bf 198} (1998) 689; K. Furuuchi, Prog. Theor. Phys. 
{\bf 103} (2000) 1043; D. H. Correa, G. Lozano, E. F. Moreno and F. A. Schaposnik, 
Phys. Lett. {\bf B515} (2001) 206; 
C. Chu, V. V. Khoze and G. Travaglini, Nucl. Phys. {\bf B621} (2002) 101.
\bibitem{mono}S. Moriyama, Phys. Lett. {\bf B485} (2000) 278; D. J. Gross and N. A. Nekrasov, 
JHEP {\bf 0007} (2000) 034.
\bibitem{gi} J. A. Harvey, hep-th/0105242; C. Sochichiu, hep-th/0202014.
\bibitem{ss}S. Gubser and S. L. Sondhi, Nucl.Phys. {\bf B605} (2001) 395;
E. T. Akhmedov, P. De Boer, G. W. Semenoff, Phys. Rev. {\bf D64} (2001) 065005; S. Sarkar and
B. Sathiapalan, JHEP {\bf 0105}, 049 (2001) 049; S. Sarkar, JHEP {\bf 0206} (2002) 003;  
L. Griguolo and M. Pietroni, JHEP {\bf 0105} (2001) 032.
\bibitem{mdnn} M. R. Douglas and N. A. Nekrasov, Rev. Mod. Phys. {\bf 73} (2001) 977.
\bibitem{grosse} H. Grosse, C. Klimcik and P. Presnajder, Int. J. Theor. Phys. {\bf 35}, (1996) 231; H. Grosse and 
A. Strohmaier, Lett. Math. Phys. {\bf 48} (1999) 163.
\bibitem{pres} P. Presnajder, J. Math. Phys. {\bf 41} (2000) 2789, H. Akoi, S. Iso, K.Nagao, hep-th/0209137.
\bibitem{madore} J. Madore, Class. Quant. Grav. {\bf 9} (1992) 69.
\bibitem{sb}H. Grosse, C. Klimcik and P. Presnajder, Commun. Math. Phys. {\bf 1788} (1996) 507;
 S. Baez, A. P. Balachandran, S. Vaidya and B. Ydri, Commun. Math. Phys. {\bf 208} (2000) 787;
A. P. Balachandran, X. Martin and D. O'Connor, Int. J. Mod. Phys. {\bf A16} (2001) 2577;
S. Vaidya, JHEP {\bf 0201} (2002) 011.
\bibitem{sv}A. P. Balachandran and S. Vaidya, Int. J. Mod. Phys. {\bf A16} (2001) 17.
\bibitem{apb} A. P. Balachandran, T. R. Govindarajan and B. Ydri, Mod. Phys. Lett. {\bf A15} (2000) 1279; 
A. P. Balachandran, T. R. Govindarajan and B. Ydri, hep-th/0006216; B. Ydri, Fuzzy Physics, PhD thesis (Syracuse Uni), 
hep-th/0110006.
\bibitem{wata} C. Klimcik, Commun. Math. Phys. {\bf 199} (1998) 257; U. C-Watamura and S. Watamura, Commun. Math. Phys. 
{\bf 212} (2000) 395; S. iso, Y. Kimura, K. Tanaka and K. Wakatsuki, Nucl. Phys. {\bf B604}, (2001) 121.
\bibitem{uv1} S. Vaidya, Phys. Lett. {\bf B512} (2001) 403; C-S Chu, J. Madore and H. Steinacker, JHEP {\bf 0108} 
(2001) 038.
\bibitem{gkp}H. Grosse, C. Klimcik and P. Presnajder, Commn. Math. Phys. {\bf 178} (1996) 507.
\bibitem{ccy} C-T Chan, C-M Chen, H. S. Yang, hep-th/0106269.
\bibitem{grs}H. Grosse and C. W. Rupp, math-ph/0103003;  H. Grosse, C. W. Rupp and A. Strohmaier, 
J. Geom. Phys. {\bf 42} (2002) 54.
\bibitem{cpn} B-H. Lee, K. Lee and H. S. Yang, Phys. Lett. {\bf B498} (2001) 277; C-T. Chan, C-M. Chen, F-L. Lin and 
H. S. Yang, Nucl. Phys. {\bf B625} (2002) 327.
\bibitem{susycpn} H. O. Girotti, M. Gomes, V. O. Rivelles and A. J. da Silva, Int. J. Mod. Phys. {\bf A17} (2002) 1503; 
H. O. Girotti, M. Gomes, A. Yu. Petrov, V. O. Rivelles and A. J. da Silva, Phys. Lett. {\bf B521} (2001) 119.
\bibitem{bp} A. A. Belavin and A. M. Polyakov, JETP Lett. {\bf 22} (1975) 245.
\bibitem{adda} H. Eichenherr, Nucl. Phys. {\bf B146} (1978) 215; V. Golo and A. M. Perelemov, 
Phys. Lett. {\bf 79B} (1978) 
112; E. Cremmer and J. Scherck, Phys. Lett. {\bf 74B} (1978) 341; D'Adda, 
M. L\"uscher and P. Di Vecchia, Nucl.
Phys. {\bf B146} (1978) 63, ibid {\bf B152} (1979) 125.
\bibitem{wz} F. Wilczek and A. Zee, Phys. Rev. Lett. {\bf 50} (1983) 2250; 
Y-S Wu and A. Zee, Phys. Lett. {\bf 147B} (1984) 325.
\bibitem{wij} M. J. Bowick, D. Karabali and L. C. R. Wijewardhana, Nucl. Phys. {\bf B271} (1986) 417; 
A. Kovner, Phys. Lett. {\bf 224B} (1989) 299.
\bibitem{frad} E. Fradkin, 'Field theories of Condensed Matter Systems', Addison- Wesley (1991).
\bibitem{polmpl} A. M. Polyakov, Mod. Phys. Lett. {\bf A3} (1988) 325; 
A. Yu. Alekseev and S. L. Shatashvili, Mod. Phys. Lett. {\bf A3} (19988) 1551.
\bibitem{psm}  R. Shankar and M. Sivakumar, Mod. Phys. Lett. {\bf A6} (1991) 2379; 
T. R . Govindarajan, N. Shaji, R. Shankar and M. Sivakumar, Phys. Rev. Lett. {\bf 69} (1992) 721; ibid, Int. J. Mod. Phys. 
{\bf A8} (1993) 3965; S. K. Paul, R. Shankar and M. Sivakumar, Mod. Phys. Lett. {\bf A6} (1991) 553. 
\bibitem{hp} T. Holstein and H. Primakoff, Phys. Rev. {\bf 58} (1940) 1098.
\bibitem{su2q}E. K. Sklyanin, Funct. Anal. Appl. {\bf 16} (1982) 262; P. P. Kulish and N. Y. Reshetikhin, J. Sov. Math. 
{\bf 23} (1983) 2435.
\bibitem{bied} L. C. Biedenharn, J. Phys. {\bf A}: Math. Gen. {\bf 22}, (1989) L879; A. J. Macfarlane, J. Phys. {\bf A}: 
Math. Gen. {\bf 22} (1989) 4581.
\bibitem{zachos} T. Curtright, D. Fairlie and C. Zachos (Ed), Quantum Groups, (Proceedings of the Argonne Workshop, 
World Scientific),(1991); Also see, Yu Manin, Quantum Groups and Non-Commutative Geometry,
Montreal:~Les Publications CRM (1988).
\bibitem{apbgi} A. P. Balachandran and G. Immirzi, Fuzzy Nambu-Goldstone Physics, hep-th/0212133.
\end{thebibliography}
\end{document}